\documentclass[useAMS,usenatbib,letterpaper]{mn2e}
\usepackage[totalwidth=480pt,totalheight=680pt]{geometry}
\usepackage{graphicx,amssymb}
\input psfig

\title[Planet-planet scattering alone cannot explain free-floating planets]{Planet-planet scattering alone cannot explain the free-floating planet population}
\author[Veras \& Raymond]{Dimitri Veras$^{1}$\thanks{E-mail: veras@ast.cam.ac.uk}, Sean N. Raymond$^{2,3}$\thanks{E-mail: rayray.sean@gmail.com}\\
$^{1}$Institute of Astronomy, University of Cambridge, Madingley Road, Cambridge CB3 0HA
\\
$^{2}$CNRS, UMR 5804, Laboratoire d'Astrophysique de Bordeaux, 2 rue de l'Observatoire, BP 89, F-33271 Floirac Cedex, France
\\
$^{3}$Universit{\'e} de Bordeaux, Observatoire Aquitain des Sciences de l'Univers, 2 rue de l'Observatoire, BP 89, F-33271 Floirac Cedex, France}

\begin{document}

\date{Accepted 2012 January 08. Received 2012 January 05; in original form 2011 November 25}

\pagerange{\pageref{firstpage}--\pageref{lastpage}} \pubyear{2012} 

\maketitle

\label{firstpage}

\begin{abstract}
Recent gravitational microlensing observations predict a vast population of free-floating giant planets that outnumbers main sequence stars almost twofold.  A frequently-invoked mechanism for generating this population is a dynamical instability that incites planet-planet scattering and the ejection of one or more planets in isolated main sequence planetary systems.  Here, we demonstrate that this process alone probably cannot represent the sole source of these galactic wanderers.  By using straightforward quantitative arguments and N-body simulations, we argue that the observed number of exoplanets exceeds the plausible number of ejected planets per system from scattering.  Thus, other potential sources of free-floaters, such as planetary stripping in stellar clusters and post-main-sequence ejection, must be considered.
\end{abstract}


\section{Introduction}

One possible explanation for the existence of free-floating planets \citep{lucroc2000,zapetal2000,zapetal2002,bihetal2009,sumi11} is that they formed in protoplanetary disks around young stars, in systems of multiple planets.  These planetary systems subsequently underwent large-scale dynamical instabilities involving close encounters between planets and strong planet-planet scattering events that ejected some planets and left the surviving planets on perturbed orbits~\citep{rasio96,weidenschilling96,linida97,papaloizou01,ford01,ford03,marzari02}. The planet-planet scattering model can reproduce a number of properties of the observed population of extra-solar planets: its broad eccentricity distribution~\citep{adams03,veras06,chatterjee08,juric08,ford08,raymond09b,raymond10}, the distribution of orbital separations between adjacent two-planet pairs~\citep{raymond09a}, and perhaps certain resonant systems~\citep{raymond08b}.  

In order for planet-planet scattering to create the free-floating planet population, the following equation:

\begin{equation}
\frac{N_{free}}{N_{stars}} = f_{giant} \times f_{unstable} \times n_{ejec},
\label{eq1}
\end{equation}

\noindent{must} hold, where $N_{free}/N_{stars}$ is the observed frequency of free-floating giant planets of $1.8^{+1.7}_{-0.8}$ per main-sequence star~\citep{sumi11}, $f_{giant}$ is the fraction of stars with giant planets, $f_{unstable}$ is the fraction of giant planet systems that become unstable, and $n_{ejec}$ is the mean number of planets that are ejected during a dynamical instability. The terms on the right-hand side of Eq. (\ref{eq1}) are all dependent on stellar mass.  We discuss these correlations extensively in Section 3; see also \cite{kenken2008}. 

Exoplanet observations constrain the fraction of stars with gas giant planets to be larger than $\sim14$\%~\citep{cumming08,howard10,mayor11} and perhaps as high as 50\%~\citep{gould10}.  The majority of giant planets are located beyond 1 AU~\citep{butler06,udry07}\footnote{See http://exoplanet.eu/ and http://exoplanets.org/}, and their abundance increases strongly with orbital distance within the observational capabilities ($\sim5$ AU) of radial velocity surveys \citep{mayor11}.  

Given the difficulty of measuring eccentricities with radial velocity measurements~\citep{shen08,zakamska10}, the fraction of planetary systems that becomes unstable is modestly constrained by the eccentricities of surviving planets.  In addition, there is a clear positive mass-eccentricity correlation: more massive exoplanets have higher eccentricities~\citep{ribas07,ford08,wright09}.  The simplest way to reproduce the observed distributions is if a large fraction of systems -- at least 50\% but more probably 75\% or more -- become unstable, and if the giant planets' masses within systems with high-mass ($M \gtrsim M_J$) planets are roughly equal~\citep[][]{raymond10}.  The typical number of planets ejected per unstable system, $n_{ejec}$, must be an increasing function of the number of planets that form in a given system.  However, $n_{ejec}$ has been addressed only tangentially in previous studies, and we quantify this value in a consistent manner here.  

Assuming observationally motivated constraints -- $N_{free}/N_{stars} = 1.8$, $f_{giant} = 0.2$, and $f_{unstable} = 0.75$ -- each instability must eject 12 Jupiter-mass planets (i.e., $n_{ejec}=12$).  For the full range of plausible constraints -- $N_{free}/N_{stars} = 1-3.5$, $f_{giant} = 0.14-0.5$, and $f_{unstable} = 0.5-1$ -- the range of values for $n_{ejec}$ is between 2 and 50.  At first glance, these values appear implausibly large.  And as we will show, $n_{ejec}$ is indeed so large as to be inconsistent with observational constraints, which means that the planet-planet scattering model, at least in its simplest form, cannot explain the free-floating planet population.  

Section 2 is dedicated to placing theoretical constraints on $n_{ejec}$ using N-body simulations of planet-planet scattering.  We chose initial conditions that we expect to be the most efficient at ejecting planets and tested systems initialized with up to 50 planets.  Section 3 incorporates this result into our main argument and discusses additional considerations, such as the contribution from previous generations of main sequence stars which are now stellar remnants, and other potential sources of free-floaters.

\begin{figure*}
\centerline{
\psfig{figure=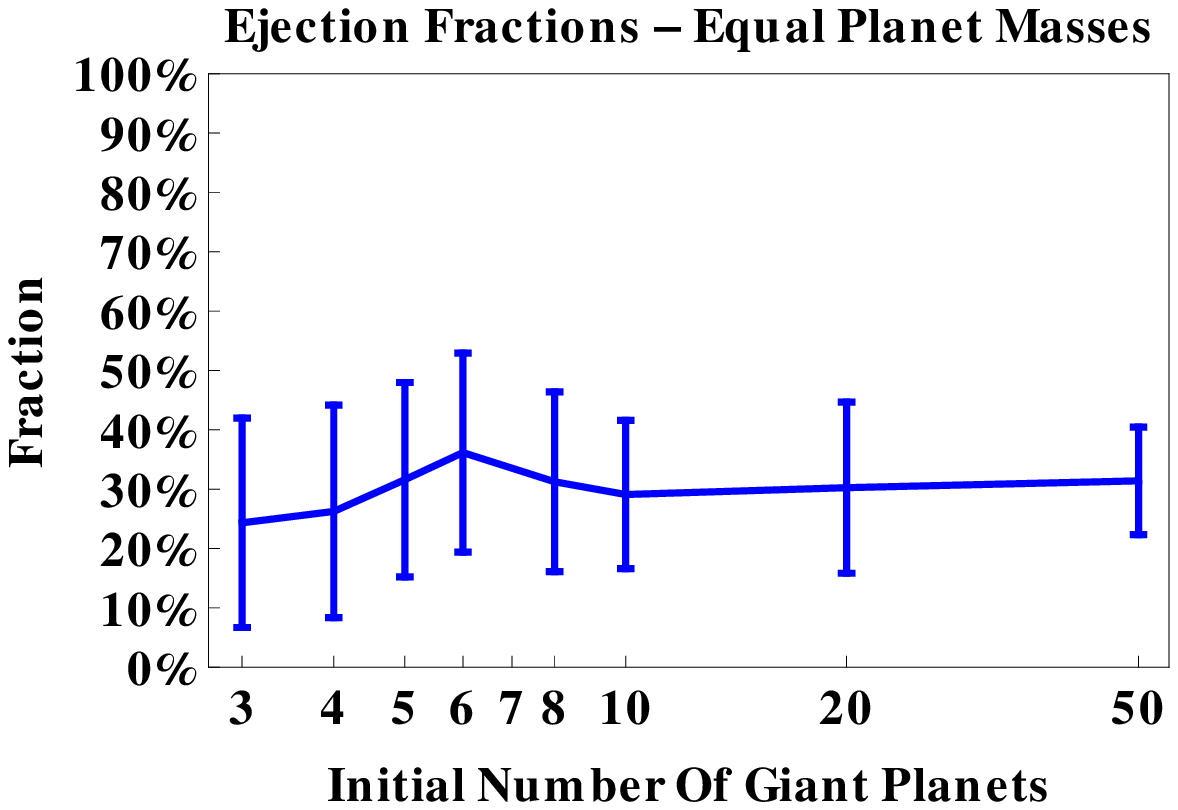,height=6cm,width=8.5cm} 
\psfig{figure=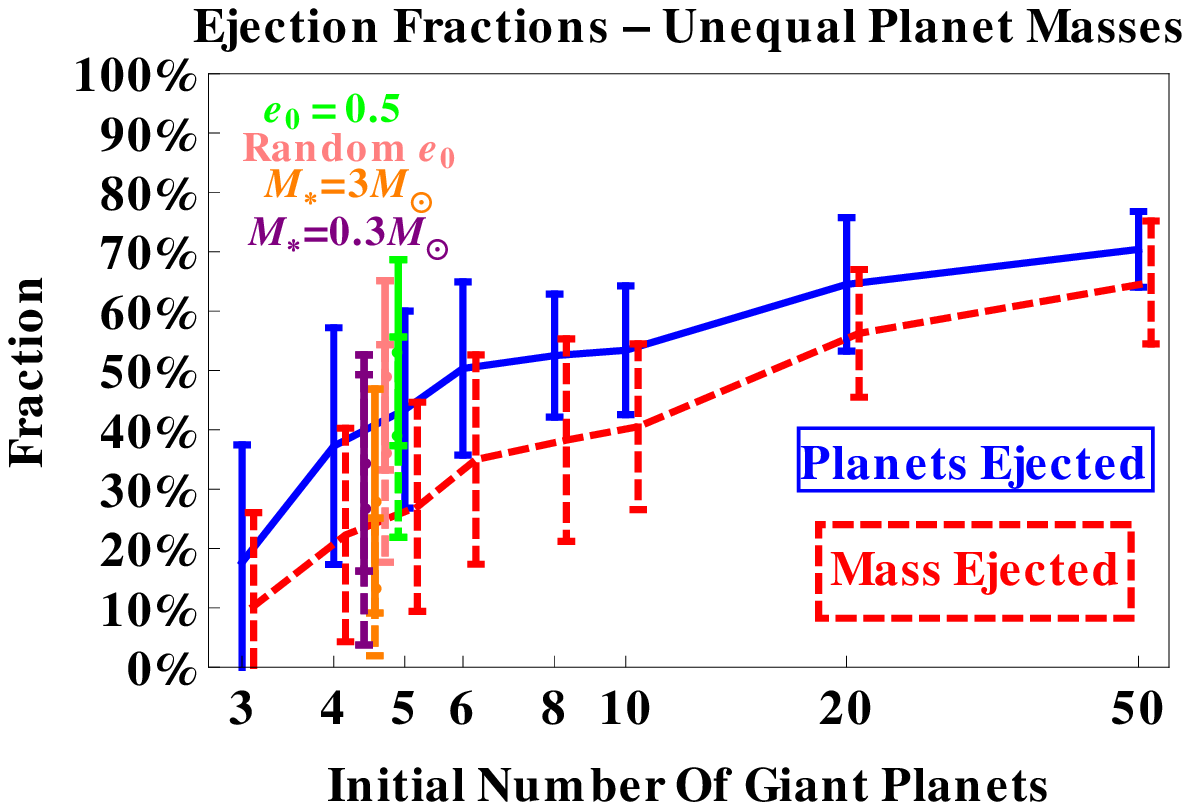,height=6cm,width=8.5cm}
}
\caption{
Ejection fractions for planetary systems with different initial numbers of giant planets ($N_p$).  The left panel shows the results for the {\tt equal} simulations, in which all planets have the same mass ($1$ Jupiter-mass).  The right panel shows the results of the {\tt random} simulations, where the planetary masses were selected randomly and logarithmically between 1 Saturn-mass and $10$ Jupiter-masses.  Vertical bars about the mean represent one standard deviation.  The solid blue curves indicate the fraction of planets that is ejected, and the red dashed curve (which is slightly offset to the right from the blue curve in the right panel for clarity) represents the fraction of initial planetary mass that is lost.  The four other segments in the right panel all represent different initial distributions for $N_p = 5$ (green: each planet has $e_0 = 0.5$; pink: each planet's initial eccentricity is assigned a random value; orange: $M_{\star} = 3 M_{\odot}$ and $a_{innermost} = 30$ AU; purple:  $M_{\star} = (1/3) M_{\odot}$ and $a_{innermost} = 0.3$ AU).  These plots provide relations between $N_p$ and $n_{ejec}$, the number of planets ejected; values from the unequal mass plot are used as coefficients in Eq. (\ref{eq2}).
}
\label{fig1}
\end{figure*}


\section{Scattering simulations}
The dynamical evolution of unstable multi-planet systems leads to planet-planet collisions, planet-star collisions, and/or the hyperbolic ejection of one or more planets.  Previous work has shown that ejections comprise the most frequent outcome~\citep{weidenschilling96,papaloizou01,marzari02,adams03,juric08,veretal2009,raymond10}.
In an extensive investigation of planet-planet scattering in systems of 10 and 50 planets, ~\cite{juric08} found that $50\%-60\%$ of all planets were ejected from unstable systems with little dependence on the initial planetary distribution.  
In addition, \cite{raymond10} showed that the number of ejected planets is larger in systems that start with unequal-mass planets and in systems with higher-mass planets.  

In order to self-consistently assess $n_{\rm ejec}$ as a function of the number of planets which have formed in a planetary system, $N_p$, we performed a suite of planet-planet scattering simulations.  Although the phase space of initial conditions is extensive, the results of the investigations referenced above (particularly from \citealt*{juric08}) demonstrate that $n_{\rm ejec}$ is largely insensitive to the initial distribution of planetary eccentricity or inclination.  Therefore, we consider initially circular and nearly-coplanar ($i$ chosen randomly from $0^\circ-1^\circ$) sets of $N_p = 3-50$ planets.  We tested two different planetary mass distributions: in the {\tt equal} simulations each planet was 1 Jupiter-mass, and in the {\tt random} simulations, the logarithm of the mass of each planet was chosen randomly such that the planet masses took values between 1 Saturn-mass and 10 Jupiter-masses.  We assumed that giant planets tend to form at a few AU -- close to the ice line -- in marginally-unstable configurations.  Thus, the innermost planet was placed at 3 AU and for $N_p =$ 3, 4, 5, 6, 8, and 10, additional planets were spread out radially with a separation of $K=4$ mutual Hill radii between adjacent planets~\citep[as $K$ determines the instability timescale;][]{marzari02,chatterjee08}.  For $N_p =$ 20 and 50, $K$ was decreased to keep the outermost planet at $\lesssim 200$ AU (for $N_p =$ 20 and 50, $K$ = 2.56 and 0.99 for the {\tt equal} simulations, and $K$ = 2.00 and 0.70 for the {\tt random} simulations).  Each system was integrated for 10 Myr using the Bulirsch-Stoer integrator in the {\it Mercury} integration package~\citep{chambers99}.  Although slow, this integrator accurately models close encounters.  The radius of each planet was taken to be Jupiter's current radius, and collisions were treated as inelastic mergers.  A planet was considered to be ejected if its orbital distance exceeded $10^5$ AU, which represents a typical distance at which galactic tides can cause escape over a typical main sequence lifetime \citep{tremaine1993}.  We ensured that each simulation conserved energy and angular momentum to one part in $10^4$ or better to avoid numerical artifacts~\citep{barnes04}.  

We obtained a sufficiently large statistical measure of $n_{\rm ejec}$ by running ensembles of systems for each value of $N_p$.  We integrated 100 sets of initial conditions for each of $N_p =$ 3, 4, 5, 6, 8, and 10, and assigned a random value to each planet's orbital angles in each instance.  We did the same for $N_p =$ 20 and 50, but instead ran 40 and 10 simulations, respectively, for each due to the increased computational cost.  Each simulation was run for $10^7$ yr.  Our results are displayed in Fig. 1.  

The figure demonstrates that between 20\% and 70\% of planets are ejected in each set of  simulations.  The vertical bars attached to the mean escape percentages for each bin represent $\pm 1$ standard deviation values from the mean.  For the {\tt equal} simulations (left panel) this mean ejection fraction is roughly constant with $N_p$, and varies by at most 10\%.  In contrast, the ejection rate in the {\tt random} simulations (right panel) increases monotonically with $N_p$.  The mean ejection percentage for $N_p = 10$ is $50\%$-$60\%$, corroborating the results from the 10-planet simulations in \cite{juric08}.  The red dashed lines represent the fraction of the initial planetary mass ejected for each value of $N_p$ (slightly offset to the right of the blue curves for clarity).  This value is less than the percentage of planets ejected in all cases, demonstrating that the most massive planets tend to scatter out the smaller ones, corroborating previous results \citep[e.g.][]{ford03}.

The effect of varying the initial system configurations does not have a drastic effect on the outcome, as demonstrated in, e.g., \cite{juric08}.  In order to sample this effect for our setup, we have performed 4 additional sets of simulations, all with $N_p = 5$, and have plotted the results on Fig. 1 to the immediate left of the $N_p = 5$ blue curves.  The green curves demonstrate the ejection percentage when each planet is initialized with $e = 0.5$ and the pink curves for a randomized distribution of initial eccentricities.  The orange and purple curves assume respectively different stellar masses ($3 M_{\odot}$ and $0.3 M_{\odot}$) and different innermost semimajor axes values ($0.3$ AU and $30$ AU) in order to account for the different approximate ice line boundaries.  In all these cases, the fraction of mass ejected is less than the fraction of planets ejected, as in the nominal $N_p = 5$ cases, and the mean ejection fraction of planets varies by about $20\%$.


\section{Discussion}

We have related ${N_{free}}/{N_{stars}}$ to $n_{ejec}$ in Section 1, and have related $n_{ejec}$ to $N_p$ in Section 2 and Fig. 1.  Now, we can link the two relations, and do so through a more precise formulation of Eq. (\ref{eq1}):

\begin{eqnarray}
&&\frac{N_{free}}{N_{stars}} = \sum_{N_p = 3}^{\infty} f_{giant}^{(N_p)} \times f_{unstable}^{(N_p)} \times n_{ejec}^{(N_p)}
\nonumber
\\
&& = 0.17 \cdot 3\cdot f_{giant}^{(3)} f_{unstable}^{(3)} + 0.37 \cdot 4\cdot f_{giant}^{(4)} f_{unstable}^{(4)} 
\nonumber
\\
&&+ 0.43 \cdot 5\cdot f_{giant}^{(5)}f_{unstable}^{(5)} + 0.50 \cdot 6\cdot f_{giant}^{(6)}f_{unstable}^{(6)} 
\nonumber
\\
&&+...
\label{eq2}
\end{eqnarray}

\noindent{where the} coefficients are taken from the mean ejection fractions of the {\tt random} simulations in Fig. 1. and which is subject to the constraint

\begin{equation}
\sum_{N_p = 3}^{\infty} f_{giant}^{(N_p)} \le 1.
\label{constraint}
\end{equation}


\noindent{Now} we can simply set ${N_{free}}/{N_{stars}} = 1.8$ and take a closer look at Eqs. (\ref{eq2})-(\ref{constraint}). 

\begin{figure}
\centerline{
\psfig{figure=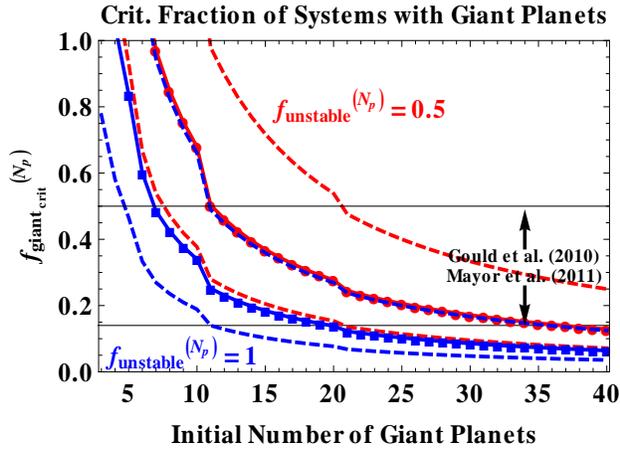,height=6cm,width=8.5cm} 
}
\caption{
The critical fraction of stellar systems that must initially contain at least $N_p$ giant planets in order to produce the free-floating planet population.  The curves were obtained from Eq. (\ref{eq2}) assuming ${N_{free}}/{N_{stars}} = 1.8^{+1.7}_{-0.8}$; the solid curves with symbols assume the nominal value of $1.8$, and the dashed curves without the symbols assume values of $1.0$ and $3.5$.  All three blue curves assume every system became unstable, and the three red curves assume that half of all systems became unstable.  Current observational limits for systems which contain {\it any} number of detected planets are given by the horizontal black lines.  This plot illustrates that an unrealistically large number of giant planets must form per star if planet-planet scattering represents the sole source of the free-floating planet population.
}
\label{fig3}
\end{figure}

   In order to satisfy the constraint in Eq. (\ref{constraint}), the vast majority of planetary systems must have $N_p \ge 5$.  If we use the observationally determined {\it upper} bound of $0.5$ \citep{gould10} on the right-hand side of Eq. (\ref{constraint}), then the vast majority of planetary systems must have $N_p \ge 7$.  If we use the lower bound of $0.14$ \citep{mayor11} -- which applies for Sun-like stars -- then the vast majority of planetary systems must have $N_p \ge 20$.  These minimum values of $N_p$ would be even higher if we instead derived Eq. (\ref{eq2}) from the {\tt equal} simulation results.  If, motivated by the monotonicity of the coefficients in Eq. (\ref{eq2}), we assume that all systems form a fixed number of planets, then we can calculate a critical giant planet frequency $f_{giant_{crit}}^{(N_p)}$ below which the free-floating population cannot be reproduced.  We plot this quantity in Fig. \ref{fig3} for two different values of $f_{unstable}^{(N_p)}$.  Here, coefficients for values of $N_p$ that were not sampled in our simulations were conservatively fixed at the coefficient of the next highest value of $N_p$ that was sampled.  For example, the coefficients of $f_{giant}^{(20)}$ and $f_{giant}^{(50)}$ are $0.65$ and $0.70$; hence, the coefficient of $f_{giant}^{(21)}$ is taken to be $0.70$.


Now that we have determined the relation between $N_p$ and $f_{giant_{crit}}^{(N_p)}$, we can assess the constraints on forming $N_p$ giant planets in a single system.  A simple first consideration is the outermost planet's semimajor axis before any scattering occurs.  For simplicity, assume $N_p$ 1-Jupiter-mass planets all orbit a $1 M_{\star}$ star, and that the planets are formed close enough to be on the verge of instability ($K=4$).  Then, with an adopted innermost semimajor axis of $3$ AU, $a_{outermost} = 3$~AU $\cdot (1.415)^{N_p-1}$.  Therefore, for $N_p > 13$, $a_{outermost} > 200$ AU.  Core accretion cannot form planets beyond $\approx 35$ AU \citep{dodetal2009}, gravitational instability has not yet been demonstrated to produce planets beyond $200$ AU \citep{boss2006}, and only in the most extreme cases may planet-disc interactions cause planets to migrate outward beyond 200 AU \citep{crietal2009}.  If giant planets were all formed by core accretion, then the $a < 35$ AU restriction implies $N_p \le 8$, implying that at least $40\%$ of systems, all with 8 giant planets, all must have become unstable in order to produce the free-floating planet population by planet-planet scattering alone.  Alternatively, if giant planets were all formed by gravitational instability, even in extended ``maximum-mass'' discs \citep{dodetal2009}, then not more than a few giant-planet-mass clumps form \citep{boley2009,boletal2010}.  

Radial velocity observations suggest that, within a few AU, giant planets are far more common around higher-mass stars \citep{johetal2007,lovmay2007,bowetal2010,johetal2010}.  However, low-mass stars outnumber higher-mass stars by a large factor \citep[][and references therein]{paretal2011}.  If $f_{giant}$ does indeed decrease as $M_{\star}$ decreases, then explaining free-floaters via planet-planet scattering proves to be more difficult simply because fewer total stars would form giant planets.  Consider, for example, the observed lower-bound frequencies of giant planets for different stellar masses deduced from radial velocity surveys for three categories of stars: M dwarfs \citep[3\%,][]{bonetal2011}, Solar-like stars: \citep[14\%,][]{mayor11}, and A-type stars \citep[20\%,][]{johetal2010}.  The total integrated value of $f_{giant}$ depends on the mass ranges adopted for these stellar types and the particular initial mass function assumed.  At minimum, the relative frequency of stars in these three bins should differ by a factor of a few (e.g., in the ratio $4$:$2$:$1$), which yields an integrated $f_{giant}$ of $\lesssim 8.5\%$.  If, however, there exists at least one order of magnitude more M dwarfs than Solar-type stars, then the integrated $f_{giant}$ must be $\lesssim 4.5\%$.  If this value is adopted as the true value of $f_{giant}$, then Fig. 2 demonstrates that giant planet systems must be extremely crowded, with at minimum 25 giant planets per system.  This value is almost certainly unrealistic given the arguments presented above and current observational constraints.  However, we note that microlensing surveys suggest that giant planets on more distant orbits may actually be very common around low-mass stars \citep{gould10}, in which case a typical giant planet system need only contain 4-10 planets (Fig. 2).  



To what degree do stars that have already evolved off the main sequence contribute to the free-floating planet population?  In order to estimate this contribution, we need only count the current population of white dwarfs, as stars that underwent supernovae or/and formed a black hole comprise a negligible fraction ($< 1\%$) of the total stellar population \citep{paretal2011}.  One of the most complete and least-biased samples of white dwarfs represents the local population (within 20 pc of the Sun), which is estimated to have a space density of $4.8 \pm 0.5 \times 10^{-3}$ pc$^{-3}$ \citep{holetal2008}.  The space density of stars in the Galactic Disc is $\approx 6 \times 10^{-1}$ pc$^{-3}$ \citep[][Pg. 3]{bintre2008}, implying that the fraction of the free-floating population which has arisen from dynamical instability soon after formation in already-dead stars is on the order of $1\%$.  This value is not great enough to change the results here unless $N_p$ is orders of magnitude higher for high-mass stars than for the lower-mass stars that dominate the current galactic population.

If the planet-planet scattering model can reproduce the free-floating planet population in a realistic framework, then the following observational predictions must hold.  First, the frequency of giant planets around low-mass stars must be high ($\sim 50\%$), in agreement with microlensing results \citep{gould10} and in disagreement with radial velocity surveys \citep{johetal2007,lovmay2007,bowetal2010,johetal2010,bonetal2011}.  This scenario could be explained if giant planets around M dwarfs are simply too distant to be sampled by current radial velocity surveys, further implying that giant planets are found at larger orbital separations around low-mass stars than Sun-like stars.  Second, systems containing a large number of giant planets and extending to large orbital radii must be abundant, at least at early times before they become unstable.  Such systems may be detectable by direct imaging of young stars \cite[e.g. HR 8799,][see also \citealt*{deletal2011}]{maretal2008,maretal2010}.  Third, the free-floating planet population of $N_{free}/N_{stars} = 1.8$ reported by \cite{sumi11} must be somewhat overestimated.  Fig. 2 shows that observations are far easier to reproduce if this value is closer to 1.  Finally, the vast majority of giant planet systems must go unstable.  This statement, in turn, implies that giant planets on highly-eccentric orbits should continue to be common and that both terrestrial planets and debris disks should be anti-correlated with eccentric giant planets \citep{veras06,raymond11}.

Other potential sources of free-floating planets exist.  One such source is dynamical ejection from multiple-star systems; only about two-thirds of all stars are single stars \citep{lada2006} so multiple-star systems may provide a comparable contribution to free-floaters.  Another potential source arises from external forces such as passing stars, galactic tides, or -- most relevant for close ($\lesssim 100$ AU) planets -- perturbations while stars are still in their birth clusters.  Indeed, studies of flybys on planetary systems \citep{malmberg07,malmberg11} indicate these perturbations may eject planets.  Recent cluster simulations have shown that planets with orbital radii of 5-30 AU are likely to disrupted by close stellar passages within their birth clusters at a rate of up to $\sim 10\%$~\citep{parqua2011}.  The cluster disruption rate is much higher than the the long-term effects from passing stars \citep[e.g.][]{weietal1987} or galactic tides \citep[e.g.][]{tremaine1993}, which are thought to cause disruption for semimajor axes of $\sim 10^5$ AU.  Free-floating Jupiter-mass objects can also be formed by collisions between high-mass protoplanetary discs \citep{linetal1998}, but extremely dense stellar conditions are needed to cause such collisions.  Other potential sources of free-floaters may arise from dynamical ejection prompted by mass loss during post-main sequence evolution.  \cite{veretal2011} show that planets at several hundred AU can be dynamically ejected due to mass loss from stars with progenitor masses greater than $\sim 2 M_{\odot}$; these planets may reach such wide distances through scattering over the main sequence lifetime of the system.  The interaction of planetary systems with localized Galactic phenomena such as the tidal streams, passage into and out of spiral arms from radial stellar migration, Lindblad resonances with the bar, and passing molecular clouds represent other, largely unexplored, potential causes for ejection.  Also, we cannot rule out the possibility that Jupiter-mass objects could simply represent the low-mass tail of the stellar initial mass function, and that free-floaters could help constrain the presence or extent of a planet-star gap in the initial mass function \citep[e.g.][]{chabrier2003,paretal2011}.


\section*{Acknowledgments}

We thank the referee for helpful comments.  S.N.R. is grateful to the IoA and DAMTP for their hospitality during his visit to Cambridge in November 2011, and acknowledges support from the CNRS's PNP and EPOV programs and the NASA Astrobiology Institute's VPL lead team.

\label{lastpage}


\begin{thebibliography}{99}


\bibitem[Adams \& Laughlin(2003)]{adams03} Adams, F.~C., \& 
Laughlin, G.\ 2003, Icarus, 163, 290 

\bibitem[Barnes \& Quinn(2004)]{barnes04} 
Barnes, R., \& Quinn, T.\ 2004, ApJ, 611, 494 

\bibitem[Bihain et al.(2009)]{bihetal2009} Bihain, G., 
Rebolo, R., Zapatero Osorio, M.~R., et al.\ 2009, A\&A, 506, 1169 

\bibitem[Binney \& Tremaine(2008)]{bintre2008} Binney, J., \& Tremaine, S.\ 2008, Galactic Dynamics: Second Edition, by James Binney and Scott Tremaine.~ISBN 978-0-691-13026-2 (HB).~Published by Princeton University Press, Princeton, NJ USA, 2008.,  

\bibitem[Boley(2009)]{boley2009} 
Boley, A.~C.\ 2009, ApJL, 695, L53 

\bibitem[Boley et al.(2010)]{boletal2010} Boley, A.~C., Hayfield, 
T., Mayer, L., \& Durisen, R.~H.\ 2010, Icarus, 207, 509 

\bibitem[Bonfils et al.(2011)]{bonetal2011} Bonfils, X., Delfosse, 
X., Udry, S., et al.\ 2011, arXiv:1111.5019 

\bibitem[Boss(2006)]{boss2006} Boss, A.~P.\ 2006, ApJL, 637, 
L137 

\bibitem[Bowler et al.(2010)]{bowetal2010} Bowler, B.~P., Johnson, 
J.~A., Marcy, G.~W., et al.\ 2010, ApJ, 709, 396 

\bibitem[Butler et al.(2006)]{butler06} Butler, R.~P., Wright, 
J.~T., Marcy, G.~W., et al.\ 2006, ApJ, 646, 505

\bibitem[Chabrier(2003)]{chabrier2003} Chabrier, G.\ 2003, 
PASP, 115, 763 

\bibitem[Chambers(1999)]{chambers99} Chambers, J.~E.\ 1999, 
MNRAS, 304, 793 

\bibitem[Chatterjee et al.(2008)]{chatterjee08} Chatterjee, S., 
Ford, E.~B., Matsumura, S., \& Rasio, F.~A.\ 2008, ApJ, 686, 580 

\bibitem[Crida et al.(2009)]{crietal2009} Crida, A., Masset, F., 
\& Morbidelli, A.\ 2009, ApJL, 705, L148

\bibitem[Cumming et al.(2008)]{cumming08} Cumming, A., Butler, 
R.~P., Marcy, G.~W., et al.\ 2008, PASP, 120, 531 

\bibitem[Delorme et al.(2011)]{deletal2011} 
Delorme, P., Lagrange, A.~M., Chauvin, G., et al.\ 2011, arXiv:1112.3008 

\bibitem[Dodson-Robinson et al.(2009)]{dodetal2009} 
Dodson-Robinson, S.~E., Veras, D., Ford, E.~B., 
\& Beichman, C.~A.\ 2009, ApJ, 707, 79 

\bibitem[Ford et al.(2001)]{ford01} Ford, E.~B., Havlickova, 
M., \& Rasio, F.~A.\ 2001, Icarus, 150, 303 

\bibitem[Ford \& Rasio(2008)]{ford08} 
Ford, E.~B., \& Rasio, F.~A.\ 2008, ApJ, 686, 621 

\bibitem[Ford et al.(2003)]{ford03} Ford, E.~B., 
Rasio, F.~A., \& Yu, K.\ 2003, Scientific Frontiers in 
Research on Extrasolar Planets, 294, 181 

\bibitem[Fregeau et al.(2006)]{freetal2006} Fregeau, J.~M., 
Chatterjee, S., \& Rasio, F.~A.\ 2006, ApJ, 640, 1086 

\bibitem[Gould et al.(2010)]{gould10} Gould, A., Dong, S., 
Gaudi, B.~S., et al.\ 2010, ApJ, 720, 1073 

\bibitem[Holberg et al.(2008)]{holetal2008} Holberg, J.~B., Sion, 
E.~M., Oswalt, T., et al.\ 2008, AJ, 135, 1225 

\bibitem[Howard et al.(2010)]{howard10} 
Howard, A.~W., Marcy, G.~W., Johnson, J.~A., 
et al.\ 2010, Science, 330, 653 

\bibitem[Johnson et al.(2007)]{johetal2007} 
Johnson, J.~A., Butler, R.~P., Marcy, G.~W., et al.\ 2007, ApJ, 670, 833 

\bibitem[Johnson et al.(2010)]{johetal2010} Johnson, J.~A., Aller, 
K.~M., Howard, A.~W., \& Crepp, J.~R.\ 2010, PASP, 122, 905 

\bibitem[Juri{\'c} \& Tremaine(2008)]{juric08} 
Juri{\'c}, M., \& Tremaine, S.\ 2008, ApJ, 686, 603 

\bibitem[Kennedy \& Kenyon(2008)]{kenken2008} 
Kennedy, G.~M., \& Kenyon, S.~J.\ 2008, ApJ, 673, 502 

\bibitem[Lada(2006)]{lada2006} 
Lada, C.~J.\ 2006, ApJL, 640, L63

\bibitem[Lin \& Ida(1997)]{linida97} 
Lin, D.~N.~C., \& Ida, S.\ 1997, ApJ, 477, 781 

\bibitem[Lin et al.(1998)]{linetal1998} Lin, D.~N.~C., Laughlin, 
G., Bodenheimer, P., \& Rozyczka, M.\ 1998, Science, 281, 2025 

\bibitem[Lovis \& Mayor(2007)]{lovmay2007} 
Lovis, C., \& Mayor, M.\ 2007, A\&A, 472, 657 

\bibitem[Lucas \& Roche(2000)]{lucroc2000} 
Lucas, P.~W., \& Roche, P.~F.\ 2000, MNRAS, 314, 858 

\bibitem[Malmberg et al.(2011)]{malmberg11} Malmberg, D., Davies, 
M.~B., \& Heggie, D.~C.\ 2011, MNRAS, 411, 859 

\bibitem[Malmberg et al.(2007)]{malmberg07} Malmberg, D., de 
Angeli, F., Davies, M.~B., et al.\ 2007, MNRAS, 378, 1207 

\bibitem[Marois et al.(2008)]{maretal2008} Marois, C., Macintosh, 
B., Barman, T., et al.\ 2008, Science, 322, 1348 

\bibitem[Marois et al.(2010)]{maretal2010} Marois, C., Zuckerman, 
B., Konopacky, Q.~M., Macintosh, B., \& Barman, T.\ 2010, Nature, 468, 1080

\bibitem[Marzari \& Weidenschilling(2002)]{marzari02} 
Marzari, F., \& Weidenschilling, S.~J.\ 2002, Icarus, 156, 570 

\bibitem[Mayor et al.(2011)]{mayor11} Mayor, M., Marmier, M., 
Lovis, C., et al.\ 2011, arXiv:1109.2497 

\bibitem[Papaloizou \& Terquem(2001)]{papaloizou01} 
Papaloizou, J.~C.~B., \& Terquem, C.\ 2001, MNRAS, 325, 221

\bibitem[Parker \& Quanz(2011)]{parqua2011} Parker, R.~J., \& Quanz, S.~P.\ 2011, MNRAS, 1760 

\bibitem[Parravano et al.(2011)]{paretal2011} Parravano, A., McKee, 
C.~F., \& Hollenbach, D.~J.\ 2011, ApJ, 726, 27 

\bibitem[Rasio \& Ford(1996)]{rasio96} 
Rasio, F.~A., \& Ford, E.~B.\ 1996, Science, 274, 954 

\bibitem[Raymond et al.(2011)]{raymond11} Raymond, S.~N., 
Armitage, P.~J., Moro-Mart{\'{\i}}n, A., et al.\ 2011, A\&A, 530, A62 

\bibitem[Raymond et al.(2009)]{raymond09b} Raymond, S.~N., 
Armitage, P.~J., \& Gorelick, N.\ 2009, ApJL, 699, L88 

\bibitem[Raymond et al.(2010)]{raymond10} Raymond, S.~N., 
Armitage, P.~J., \& Gorelick, N.\ 2010, ApJ, 711, 772 

\bibitem[Raymond et al.(2008)]{raymond08b} Raymond, S.~N., Barnes, 
R., Armitage, P.~J., \& Gorelick, N.\ 2008, ApJL, 687, L107 

\bibitem[Raymond et al.(2009)]{raymond09a} Raymond, S.~N., Barnes, 
R., Veras, D., et al.\ 2009, ApJL, 696, L98 

\bibitem[Ribas \& Miralda-Escud{\'e}(2007)]{ribas07} 
Ribas, I., \& Miralda-Escud{\'e}, J.\ 2007, A\&A, 464, 779 

\bibitem[Shen \& Turner(2008)]{shen08} 
Shen, Y., \& Turner, E.~L.\ 2008, ApJ, 685, 553 

\bibitem[Sumi et al.(2011)]{sumi11} Sumi, T., Kamiya, K., 
Bennett, D.~P., et al.\ 2011, Nature, 473, 349 

\bibitem[Tremaine(1993)]{tremaine1993} Tremaine, S.\ 1993, Planets 
Around Pulsars, 36, 335 

\bibitem[Udry \& Santos(2007)]{udry07} Udry, S., \& Santos, N.~C.\ 2007, AR\&A, 45, 397 

\bibitem[Veras \& Armitage(2006)]{veras06} 
Veras, D., \& Armitage, P.~J.\ 2006, ApJ, 645, 1509 

\bibitem[Veras et al.(2009)]{veretal2009} Veras, D., Crepp, J.~R., 
\& Ford, E.~B.\ 2009, ApJ, 696, 1600 

\bibitem[Veras et al.(2011)]{veretal2011} Veras, D., Wyatt, M.~C., 
Mustill, A.~J., Bonsor, A., \& Eldridge, J.~J.\ 2011, MNRAS, 417, 2104 

\bibitem[Weidenschilling \& Marzari(1996)]{weidenschilling96} 
Weidenschilling, S.~J., \& Marzari, F.\ 1996, Nature, 384, 619 

\bibitem[Weinberg et al.(1987)]{weietal1987} Weinberg, M.~D., 
Shapiro, S.~L., \& Wasserman, I.\ 1987, ApJ, 312, 367 

\bibitem[Wright et al.(2009)]{wright09} Wright, J.~T., Upadhyay, 
S., Marcy, G.~W., et al.\ 2009, ApJ, 693, 1084 

\bibitem[Zakamska et al.(2011)]{zakamska10} Zakamska, N.~L., Pan, 
M., \& Ford, E.~B.\ 2011, MNRAS, 410, 1895 

\bibitem[Zapatero Osorio et al.(2000)]{zapetal2000} Zapatero 
Osorio, M.~R., B{\'e}jar, V.~J.~S., Mart{\'{\i}}n, E.~L., et al.\ 2000, 
Science, 290, 103 

\bibitem[Zapatero Osorio et al.(2002)]{zapetal2002} Zapatero 
Osorio, M.~R., B{\'e}jar, V.~J.~S., Mart{\'{\i}}n, E.~L., et al.\ 2002, 
ApJ, 578, 536 

\end{thebibliography}
\end{document}